# Evidence and Elimination: A Bayesian Interpretation of Falsification in Scientific Practice

Tommaso Costa[1,2,3]

[1] GCS-fMRI, Koelliker Hospital and Department of Psychology, University of Turin, Turin, Italy.

[2] FOCUS Laboratory, Department of Psychology, University of Turin, Turin, Italy.

[3] Neuroscience Institute of Turin (NIT), Turin, Italy.

e-mail address: tommaso.costa@unito.it

**Abstract**

The classical conception of falsification presents scientific theories as entities that are decisively refuted when their predictions fail. This picture has long been challenged by both philosophical analysis and scientific practice, yet the relationship between Popper's eliminative view of theory testing and Bayesian model comparison remains insufficiently articulated. This paper develops a unified account in which falsification is reinterpreted as a Bayesian process of model elimination. A theory is not rejected because it contradicts an observation in a logical sense; it is eliminated because it assigns vanishing integrated probability to the data in comparison with an alternative model. This reinterpretation resolves the difficulties raised by the Duhem–Quine thesis, clarifies the status of auxiliary hypotheses, and explains why ad hoc modifications reduce rather than increase theoretical credibility. The analysis is illustrated through two classical episodes in celestial mechanics, the discovery of Neptune and the anomalous precession of Mercury. In the Neptune case, an auxiliary hypothesis internal to Newtonian gravity dramatically increases the marginal likelihood of the theory, preserving it from apparent refutation. In the Mercury case, no permissible auxiliary modification can rescue the Newtonian model, while general relativity assigns high probability to the anomaly without adjustable parameters. The resulting posterior collapse provides a quantitative realisation of Popper's eliminative criterion. Bayesian model comparison therefore supplies the mathematical structure that Popper's philosophy lacked and offers a coherent account of scientific theory change as a process of successive eliminations within a space of competing models.

# 1. Introduction

The traditional view of scientific rationality, strongly influenced by the early interpretation of Karl Popper's falsificationism, holds that theories are tested through attempts to refute them in light of empirical evidence. According to this picture, scientific progress occurs when a theory fails a severe test and is therefore rejected. However, Popper's more mature writings, especially in *Conjectures and Refutations* (1963) and *Objective Knowledge* (1972), make it clear that this is not an accurate representation of how theory change actually works in science. Popper repeatedly insisted that no theory can ever be logically falsified by a single observation, that anomalies are open to reinterpretation, and that the rejection of a theory is meaningful only in the presence of a better rival. What Popper called falsification is therefore better understood as a comparative and eliminative process rather than as a logical contradiction between theory and observation.

A complementary tradition within the foundations of scientific inference, developed primarily by Harold Jeffreys (1939), I. J. Good (1950, 1960, 1983) and, later, Edwin T. Jaynes (2003), argues that rational theory choice is inherently Bayesian. In this view, the strength of evidence for one hypothesis relative to another is determined by their integrated likelihoods and by the posterior odds that follow from Bayes' theorem. From this perspective, no hypothesis is ever refuted in isolation; it is simply rendered less plausible than its competitors. Jeffreys saw the Bayes factor as the fundamental tool for scientific inference and argued that the problem of induction disappears once one recognises that the central task of science is to compare hypotheses, not to validate or falsify them individually.

Despite their conceptual affinity, the connection between Popper's eliminative conception of scientific rationality and the Bayesian theory of model comparison has rarely been made explicit. Popper's writings emphasise bold conjectures, severe tests and the elimination of theories through criticism, but do not provide a quantitative measure of what makes one theory more severely tested than another. Conversely, the Bayesian literature has developed a rich calculus of evidence, explicativity and posterior support, but has seldom been related to the philosophical tradition of critical rationalism. As a result, the two frameworks have often been presented as mutually incompatible, even though their underlying logic is remarkably similar.

This paper aims to bring these traditions together. The central claim is that Popperian falsification can be coherently understood only when reformulated as Bayesian model elimination. Under this interpretation, the problem of theory rejection reduces to a comparison of posterior plausibilities. A theory is rationally rejected when its posterior probability becomes negligibly small relative to that of its alternatives. This reading dissolves many of the classical objections to falsificationism, including the problem of auxiliary hypotheses and the Duhem–Quine thesis, because it makes explicit that hypotheses are never tested in isolation but only as part of a space of rival models.

To illustrate this claim, the paper develops two detailed case studies drawn from classical celestial mechanics. The discovery of Neptune is presented as a clear example of a Popperian situation in which an apparent anomaly does not refute a theory because there exists a compelling rival explanation within the same theoretical framework. Jaynes' Bayesian reconstruction of this episode shows that the Newtonian prediction remains overwhelmingly plausible once one incorporates the hypothesis of an unseen planet. By contrast, the anomalous precession of Mercury's perihelion provides an example of eliminative falsification in action. Here, the Newtonian framework cannot accommodate the data even when augmented with auxiliary hypotheses such as a hypothetical planet inside Mercury's orbit, and the posterior odds fall decisively in favour of Einstein's general relativity.

The upshot is that Popperian falsification, once purified from the myth of instant refutation, is entirely compatible with a Bayesian understanding of scientific inference. Scientific progress emerges not from the logical defeat of isolated hypotheses but from a quantitative process of elimination in which theories are compared, tested and replaced by those with greater evidential support. This unified interpretation clarifies the structure of theory change and suggests that Bayesian model comparison provides the natural mathematical completion of Popper's critical rationalism.

## 2. Historical Background

The history of the foundations of scientific reasoning presents a fragmented landscape in which several authors approached, from different angles, the problem of how theories are tested, refined and replaced. Although their vocabularies and motivations differ, many of these thinkers converged on the idea that scientific rationality is inherently comparative: theories are not validated in isolation, but assessed relative to alternatives. What is largely absent in the historical literature is an explicit recognition that this comparative structure is naturally formalised by the Bayesian theory of model comparison. To clarify this connection, it is useful to review the contributions of Popper, Jeffreys, Good, Jaynes and Lakatos, as each of them captured a crucial aspect of eliminative reasoning while falling short of integrating it into a unified mathematical framework.

Karl Popper's early writings, especially in *Logik der Forschung* (1934), are often interpreted as advocating an image of science in which theories face potentially decisive refutations and are discarded when predictions fail. Popper's later work, however, presents a more nuanced view. In *Conjectures and Refutations* (1963) he emphasises that scientific knowledge grows through a process of proposing bold hypotheses and subjecting them to critical scrutiny, but he also stresses that no single observation can definitively falsify a theory. Anomalies are ambiguous unless an alternative hypothesis is available to absorb the evidence. Popper's notion of falsification is therefore intrinsically comparative: a theory may be said to be refuted only when a rival theory offers a better explanation of the available evidence. Yet Popper never articulated a quantitative measure of comparative support, and his philosophical analysis remained disconnected from the statistical tools that could implement his eliminative vision.

Harold Jeffreys approached the problem from the opposite direction. In his *Theory of Probability* (1939) he sought to ground scientific inference on Bayesian principles and argued that Bayesian updating provides a general logic for comparing hypotheses in light of data. Jeffreys insisted that the fundamental unit of evidence is the Bayes factor, which measures how the data modify the odds between competing hypotheses. In his view, rejection and acceptance are always relative, and the task of scientific inference is to determine which hypothesis has the greatest integrated likelihood. Jeffreys' mathematics thus provides what Popper's falsificationism lacked: a precise calculus for evaluating the strength of evidence. Despite this, Jeffreys did not relate his Bayesian account to Popper's philosophy and did not frame model comparison as a form of eliminative induction, even though the connection is conceptually direct.

I. J. Good extended Jeffreys' work and developed the modern theory of evidential support. In a long series of papers and monographs, including *Probability and the Weighing of Evidence* (1950) and *Good Thinking* (1983), he introduced concepts such as the weight of evidence, the Bayes factor and explicativity, which he described as a measure of the ability of a hypothesis to explain a structured pattern of observations. Good's analysis is deeply eliminative in spirit: hypotheses are never judged in isolation but always in comparison with rivals, and evidence serves to redistribute plausibility across a space of alternative models. Good hinted that this approach captures much of what Popper intended, but he did not explicitly develop a connection with falsificationism or with the philosophical literature on theory change.

Edwin T. Jaynes provided perhaps the most compelling demonstration of the Bayesian view of scientific reasoning. In *Probability Theory: The Logic of Science* (2003), published posthumously, Jaynes reconstructs classical scientific episodes—such as the discovery of Neptune and the anomalous precession of Mercury—to show how Bayesian reasoning accounts for the acceptance and rejection of theories. Jaynes argues that anomalies do not automatically threaten a theory; rather, their significance depends on whether alternative explanations provide a more plausible account of the data. In the Uranus–Neptune case, an auxiliary hypothesis internal to Newtonian mechanics yields a dramatic improvement in plausibility, thereby preserving the core theory. In the Mercury case, no such auxiliary hypothesis succeeds, and general relativity emerges with overwhelming posterior support. Jaynes did not frame this analysis in Popperian terms, but his narrative reveals that the eliminative logic Popper described qualitatively is precisely what Bayesian model comparison captures quantitatively.

Imre Lakatos attempted to reconcile Popperian falsificationism with the historical development of science. In his theory of research programmes, presented in *The Methodology of Scientific Research Programmes* (1978), Lakatos argued that theories are embedded in evolving frameworks with hard cores and protective belts of auxiliary assumptions. A programme is progressive when its modifications lead to novel predictions that are confirmed, and degenerative when adjustments are introduced merely to shield the core theory from anomalies. Lakatos clearly recognised that theory assessment is comparative and that anomalies matter only when alternative programmes produce better explanations. Yet Lakatos did not offer a mathematical account of progressiveness or degeneracy, nor did he connect his ideas to the Bayesian literature that had already formalised comparative inference.

Across these diverse traditions, a common structure emerges. Popper identified the eliminative nature of scientific criticism but lacked a quantitative theory to support it. Jeffreys and Good developed the quantitative theory but did not relate it to the philosophical discourse on falsification. Jaynes demonstrated the power of Bayesian reasoning in scientific practice without explicitly linking it to Popper or Lakatos. Lakatos captured the dynamics of theory competition but without a calculus of evidential support. The central thesis of this paper is that these strands can be integrated into a single framework in which falsification is understood as a Bayesian process of model elimination. Under this view, scientific theories are not refuted by isolated observations but are rendered obsolete when their posterior support collapses relative to that of superior alternatives.

## 3. Bayesian Model Comparison as Eliminative Inference

The Bayesian theory of model comparison provides a transparent and mathematically coherent account of how evidence redistributes plausibility across competing hypotheses. In this framework, a scientific theory is represented as a statistical model that assigns probabilities to observable outcomes. When data become available, the posterior probability of each model is obtained by combining its prior probability with its marginal likelihood, the latter being the probability of the data under that model, averaged over its parameter space. The central quantity governing this redistribution is the Bayes factor, introduced formally by Jeffreys (1939) and later popularised by Good (1950), which expresses the support that the data provide for one model relative to another.

Posterior probabilities arise from Bayes' theorem applied at the level of models. If $M_1$ and $M_2$ are competing theories and $D$ is the observed data, then the posterior odds are equal to the prior odds multiplied by the Bayes factor. This simple identity captures the essence of eliminative reasoning. A model is never rejected in isolation; it is eliminated only to the extent that another model receives greater support. Posterior probabilities therefore provide a graded measure of theoretical survival, in stark contrast to the all-or-nothing conception of refutation that is often attributed to Popper's early work. The Bayesian framework reveals that the very notion of refutation becomes meaningful only when a theory's posterior probability becomes negligible relative to its competitors.

This interpretation dissolves the classical problem, highlighted by Duhem and Quine, that hypotheses are never tested in isolation because empirical predictions depend on auxiliary assumptions. In a Bayesian analysis, auxiliary assumptions are simply additional components of the model. A hypothesis together with its auxiliaries is treated as a single composite model whose plausibility is assessed relative to alternative composite models. If an anomaly can be accommodated by modifying or extending the auxiliaries in a way that improves overall plausibility, the posterior probability of the composite model will increase rather than decrease. Conversely, if attempted modifications introduce penalties due to increased complexity or poor predictive performance, the marginal likelihood of the model will fall, leading to its elimination relative to simpler or more predictive rivals. The Bayesian framework therefore captures, in quantitative form, the insight expressed by Lakatos that a research programme is undermined not by isolated counterexamples but by the comparative degeneracy of its theoretical adjustments.

The marginal likelihood plays a central role in this process because it naturally penalises ad hoc hypotheses. Since the marginal likelihood averages the likelihood of the data over the entire parameter space, a model that introduces additional parameters without improving predictive accuracy will assign the data a lower probability than a more parsimonious model. Jaynes repeatedly emphasised that this is the mathematical expression of Occam's razor. A theory that fits the data only after fine-tuning many adjustable parameters will have a lower integrated likelihood than a theory that predicts the same data more directly, even if both achieve similar goodness-of-fit at their optimum parameter values. In Bayesian terms, such a theory has less explicativity in the sense discussed by Good and Jaynes: it provides a narrower, more contrived account of the observed structure.

These considerations reveal why Bayesian model comparison is naturally eliminative. Theories compete for posterior probability mass, and those with poor predictive performance, excessive complexity or weak integrated support gradually lose plausibility. A theory is eliminated when its posterior becomes dominated by that of an alternative model. This process is continuous, cumulative and relative, precisely the structure Popper had in mind when insisting that falsification requires the existence of a superior rival. Bayesian inference does not interpret anomalies as logical contradictions but as shifts in relative plausibility, and thus provides the normative foundation that Popper's qualitative framework lacked.

Under this view, the scientific method becomes a sequence of eliminations rather than confirmations or refutations. Theories that fail to predict new observations competitively are gradually relegated to negligible posterior weight, while those that exhibit strong explicativity gain prominence. Scientific progress therefore appears as a process of successive reallocations of plausibility, a dynamic closely aligned with Popper's intention and made precise by the Bayesian calculus of evidence.

## 4. Good's Explicativity and Jeffreys' Evidence

The Bayesian framework for theory assessment did not emerge all at once. Its conceptual and mathematical foundations were shaped in stages by the contributions of Jeffreys, Good and Jaynes, each of whom emphasised a different aspect of how evidence should guide scientific inference. The cumulative result is a theoretical architecture in which the strength of a hypothesis is determined by the probability it assigns to the observable world. What distinguishes this approach from classical accounts of induction is the central role of the marginal likelihood and the associated measures of evidential weight.

Jeffreys defined the fundamental quantity for comparing scientific theories as the integrated likelihood of the data under a model. If $M$ is a model with parameter vector $\theta$, likelihood $p(D \mid \theta, M)$, and prior $p(\theta \mid M)$, the marginal likelihood is

$$p(D \mid M) = \int p(D \mid \theta, M) \, p(\theta \mid M) \, d\theta.$$

This integral expresses the model's capacity to predict data before the parameters are tuned to fit them. A model that explains data only through precise adjustments to its parameters will assign them a lower probability when averaged over the full parameter space. Jeffreys insisted that this quantity reflects the total predictive content of a model and that scientific inference should be grounded on the comparison of marginal likelihoods rather than on point estimates or significance tests. His concept of the Bayes factor, defined as

$$B_{12} = \frac{p(D \mid M_1)}{p(D \mid M_2)},$$

provided a direct measure of how evidence shifts the odds between two competing theories. Jeffreys viewed this ratio as the quantitative expression of the severity of a test, anticipating Popper's later insistence that severe tests discriminate between competing explanations.

I. J. Good expanded this line of thought by analysing the structure of evidential support in greater depth. Good introduced the notion of the weight of evidence, defined as the logarithm of the Bayes factor,

$$w_{12} = \log B_{12},$$

which has the appealing property of additivity across independent data. Good repeatedly emphasised that evidence does not accumulate in favour of a theory in isolation but always relative to its competitors. From this observation he developed the idea of explicativity, the notion that a theory is preferable if it provides a high probability for the particular pattern exhibited by the data. Explicativity is therefore not a measure of goodness-of-fit in isolation but of the relationship between the complexity of a hypothesis and the structure of the observations it aims to explain. A theory that anticipates the qualitative form of the data without requiring the parameters to be finely adjusted is assigned a high explicativity and therefore tends to dominate in the posterior distribution of models.

Good's analysis revealed that the predictive structure of a model is as important as its empirical adequacy. This insight is captured mathematically by the fact that the marginal likelihood penalises models that disperse their prior mass over a wide range of parameter values without concentrating probability on the observed data. Suppose a model $M_1$ predicts a broad and diffuse range of possible datasets, while a rival model $M_2$ sharply predicts the qualitative shape of the data. Even if both models can be tuned to fit the observations, $M_2$ will have a higher marginal likelihood because it assigns a larger fraction of its prior mass to regions of the sample space consistent with the actual data. Good's explicativity is therefore a formalisation of the classical intuition that good theories explain much with little, an idea that Popper occasionally expressed and that Lakatos associated with progressive research programmes.

Jaynes synthesised and extended these contributions by arguing that Bayesian probability theory is the unique mathematical representation of rational inference under uncertainty. In his analysis of scientific reasoning, Jaynes emphasised that marginal likelihoods and Bayes factors provide the only consistent way to evaluate hypotheses in the presence of ambiguous or incomplete data. He showed that predictive accuracy must be balanced against model complexity through the integrated likelihood and that this balance is achieved automatically by the rules of probability. Jaynes openly criticised both frequentist hypothesis testing, which he believed misrepresents evidence, and the philosophical accounts of induction that rely on intuitive notions of refutation. His reconstructions of classical scientific episodes demonstrate that the success or failure of a theory depends on its ability to assign high probability to the observed facts when integrated over its full parameter space.

Jaynes' treatment of Neptune and Mercury makes this point vivid. In the Neptune case, the auxiliary hypothesis of an unobserved planet greatly increases the marginal likelihood of the Newtonian model relative to the alternative that attributes the discrepancy to measurement error. In the Mercury case, the hypothesis of an intra-Mercurial planet has negligible predictive power once observational constraints are included, causing the marginal likelihood of Newton's theory to collapse relative to that of general relativity. In both cases, the mathematical mechanism is identical: the elimination of a model follows from the fact that it assigns vanishing integrated probability to the actual data. This result echoes the eliminative logic of Popperian falsification yet is expressed within a framework that resolves the ambiguities associated with auxiliary assumptions and ad hoc modifications.

What emerges from the contributions of Jeffreys, Good and Jaynes is a coherent picture in which scientific theories survive by concentrating probability on the observed structure of the world. A theory is eliminated not because it contradicts a datum in a logical sense but because it predicts the datum poorly relative to its competitors. The Bayesian machinery quantifies this relationship, and the notion of explicativity captures the extent to which a theory illuminates the structure of the data without undue complexity. In this sense, Bayesian model comparison provides the formal apparatus necessary to turn Popper's qualitative vision of critical theory selection into a precise mathematical theory of eliminative inference.

## 5. Case Study I: Neptune and the Bayesian Preservation of Newtonian Gravity

The discovery of Neptune in 1846 is often portrayed as a triumph of Newtonian celestial mechanics, yet its deeper methodological significance lies in the way it illustrates the logic of eliminative inference. When astronomers noticed persistent deviations in the orbit of Uranus from the predictions of Newtonian theory, the phenomenon presented what could have appeared as a counterexample to the gravitational law. The question facing the scientific community was whether this anomaly represented a genuine threat to Newtonian mechanics or whether it could be absorbed by an auxiliary hypothesis internal to the theory. The episode is illuminating because it shows that the significance of an anomaly depends entirely on the comparative plausibility of alternative explanations, a point that Jaynes (2003) emphasised in his analysis of the case.

The residuals in the observed motion of Uranus, defined as the differences between observed positions and those predicted by Newtonian gravitation with all known planetary masses, were not random noise but exhibited a structured pattern. Let $y = (y_1, \ldots, y_n)$ denote the sequence of residuals and assume, following Jaynes, that they can be represented by the model

$$y_i = A g_i + \varepsilon_i,$$

where $A$ is a scalar parameter representing the amplitude of the gravitational influence of a hypothesised planet, $g_i$ is the deterministic pattern that such a planet would produce at the $i$-th observational epoch, and $\varepsilon_i$ is observational noise. For simplicity, it is assumed that $\varepsilon_i \sim \mathcal{N}(0, \sigma^2)$ independently. The vector $g = (g_1, \ldots, g_n)$ is determined by orbital perturbation theory once a candidate location for the unseen planet is specified. The hypothesis of no additional planet corresponds to the special case $A = 0$.

Bayesian model comparison between the hypothesis $H_0$ that no further planet exists and the alternative $H_1$ that there is a planet with unknown amplitude $A$ proceeds by evaluating their marginal likelihoods. Under $H_0$, the data follow

$$p(y \mid H_0) = (2\pi\sigma^2)^{-n/2} \exp\left(-\frac{y^\top y}{2\sigma^2}\right).$$

Under $H_1$, the likelihood conditional on $A$ is

$$p(y \mid A, H_1) = (2\pi\sigma^2)^{-\frac{n}{2}} \exp\left(-\frac{(y - Ag)^\top (y - Ag)}{2\sigma^2}\right).$$

Jaynes argues that the physically natural prior for $A$ is a normal distribution with mean zero and variance $\tau^2$, reflecting uncertainty about the mass and orbital position of the putative planet. With

$$A \sim \mathcal{N}(0, \tau^2),$$

the marginal likelihood under $H_1$ is obtained by integrating over $A$:

$$p(y \mid H_1) = \int_{-\infty}^{+\infty} p(y \mid A, H_1) p(A \mid H_1) \, dA.$$

A straightforward calculation, first expanding $(y - Ag)^\top (y - Ag)$, leads to

$$p(y \mid H_1) = (2\pi\sigma^2)^{-\frac{n}{2}} \exp\left(-\frac{y^\top y}{2\sigma^2}\right) \exp\left(\frac{d^2}{2c}\right) \left(\frac{1}{\tau^2 c}\right)^{1/2},$$

where

$$c = \frac{g^\top g}{\sigma^2} + \frac{1}{\tau^2}, d = \frac{g^\top y}{\sigma^2}.$$

The Bayes factor in favour of the planetary hypothesis $H_1$ over $H_0$ is therefore

$$B_{10} = \frac{p(y \mid H_1)}{p(y \mid H_0)} = \left(\frac{1}{\tau^2 c}\right)^{1/2} \exp\left(\frac{d^2}{2c}\right).$$

To express the Bayes factor in a more interpretable form, Jaynes introduces two dimensionless quantities. The first,

$$\lambda = \frac{\tau^2}{\sigma^2} g^\top g,$$

describes the information content of the data combined with the prior uncertainty on the amplitude. The second,

$$z^2 = \frac{(g^\top y)^2}{\sigma^2 g^\top g},$$

measures how strongly the data align with the signal pattern predicted by the hypothesised planet. With these definitions, one finds

$$\tau^2 c = 1 + \lambda,$$

and

$$\frac{d^2}{2c} = \frac{1}{2}\frac{\lambda}{1+\lambda} z^2,$$

yielding the compact expression

$$B_{10} = (1+\lambda)^{-1/2} \exp\left(\frac{1}{2}\frac{\lambda}{1+\lambda} z^2\right).$$

This formula reveals the structure of the inference with great clarity. When the data exhibit a pattern closely aligned with the predicted perturbation induced by a hypothesised planet, the quantity $z^2$ becomes large, and the exponential term dominates, driving the Bayes factor to extremely large values. In such a case, the alternative explanation that attributes the deviations to noise alone becomes overwhelmingly implausible. Jaynes notes that even under conservative assumptions on the prior variance $\tau^2$, the Bayes factor associated with the Uranus residuals was so large that the existence of a new planet was, in a Bayesian sense, virtually certain before its telescopic confirmation by Galle.

The Bayesian analysis therefore explains why Newtonian mechanics was not threatened by the anomaly in Uranus' orbit. There existed an auxiliary hypothesis internal to the Newtonian framework that dramatically increased the marginal likelihood of the theory. Popper's insistence that anomalies do not refute a theory unless a better alternative exists finds a natural mathematical counterpart in the fact that the posterior odds for Newtonian gravity, augmented with the planetary hypothesis, were enormously greater than those for any rival hypothesis. The discovery of Neptune thus exemplifies how a theory can survive severe empirical scrutiny through the introduction of a well-motivated auxiliary assumption that increases its explicativity without compromising its simplicity.

Jaynes' reconstruction of this episode also provides a counterexample to simplistic accounts of falsification. If scientific theories were judged solely by whether their predictions match observations, then Newton's law of gravitation should have been rejected as soon as the discrepancy in Uranus' orbit was identified. Instead, the anomaly motivated the search for an auxiliary cause, and the quantitative analysis of model plausibility shows that this search was not an ad hoc manoeuvre but a rational inference driven by the structure of the data. The survival of Newtonian gravity in the Uranus case, therefore, is not a matter of dogmatic adherence but a consequence of its overwhelming posterior probability relative to competing hypotheses. This provides the first half of the contrast that will be completed by the Mercury case in the next section.

**6. Case Study II: Mercury and the Bayesian Elimination of Newton's Theory**

The anomalous precession of Mercury's perihelion presents a historically and conceptually different challenge from the Uranus–Neptune case. Here the anomaly was not an irregular series of deviations distributed across many observations but a single, precise discrepancy: the observed precession exceeded the Newtonian prediction by approximately forty–three arcseconds per century. The Newtonian model, including all known perturbations from the other planets, yielded a residual precession that could not be eliminated by refinements of observational accuracy or adjustments to planetary masses. The crucial question is whether this anomaly could be absorbed by an auxiliary hypothesis within the Newtonian framework or whether it forced the adoption of a rival theory with superior evidential support. Jaynes (2003) argues that the Bayesian analysis resolves this question decisively.

Let $y$ denote the observed anomalous precession and let $\mu$ represent the true anomalous precession produced by whatever gravitational law governs Mercury's orbit. Measurement error can be modelled as a normal deviation with standard deviation $\sigma$, so that

$$y \mid \mu \sim \mathcal{N}(\mu, \sigma^2).$$

Observational analyses from the late nineteenth and early twentieth centuries placed the anomalous component at approximately $y = 43$ arcseconds per century, with measurement uncertainty on the order of $\sigma = 0.5$. This value is sufficiently small that the anomaly cannot plausibly be attributed to observational noise alone.

The Newtonian strategy for accommodating the discrepancy was to posit the existence of an additional planet interior to Mercury's orbit, often referred to as Vulcan, or alternatively, a ring of intra–Mercurial matter. Under this hypothesis, the anomalous precession arises because the gravitational field of the additional mass produces a perturbation in Mercury's orbit. If $\theta$ denotes the precession induced by the hypothesised planet, then the Newton–plus–Vulcan model $H_2$ implies

$$y \mid \theta, H_2 \sim \mathcal{N}(\theta, \sigma^2).$$

Prior to accounting for observational constraints, it is reasonable to model $\theta$ as a normal random variable with mean zero and variance $\tau^2$, reflecting uncertainty regarding the mass and orbital radius of the hypothetical object. Thus

$$\theta \mid H_2 \sim \mathcal{N}(0, \tau^2).$$

Marginalising over $\theta$ yields

$$y \mid H_2 \sim \mathcal{N}(0, \sigma^2 + \tau^2),$$

a distribution that describes the natural range of anomalous precessions compatible with variations in the characteristics of the intra–Mercurial object.

If we take $\tau = 20$ arcseconds per century, a deliberately generous choice that favours the Vulcan hypothesis, then $\sigma^2 + \tau^2 = 400.25$ and the standard deviation of the predictive distribution is approximately 20.006. Under this model, the probability density of observing $y = 43$ is

$$p(y \mid H_2) = \frac{1}{\sqrt{2\pi(\sigma^2 + \tau^2)}} \exp\left(-\frac{y^2}{2(\sigma^2 + \tau^2)}\right),$$

which numerically evaluates to approximately $2 \times 10^{-3}$. This value is small but not negligible. At this stage, the Newtonian model augmented with a hypothetical planet is not eliminated; it retains some credibility relative to a rival hypothesis even though the data are somewhat improbable under it.

The decisive difficulty for the Newtonian programme, however, is that the very existence of a planet capable of producing a precession of forty–three arcseconds is incompatible with systematic observational evidence accumulated over decades. Any intra–Mercurial planet of sufficient mass to affect Mercury's orbit would necessarily produce observable perturbations in solar transits or in the motion of other planets, none of which was ever detected. This additional evidential constraint can be encoded by introducing an auxiliary observation $w$ summarising the fact that no such object was observed. If the effect $\theta$ were large, then $w$ would also deviate from zero. A simple model is

$$w \mid \theta, H_2 \sim \mathcal{N}(\theta, \sigma_2^2),$$

where $w \approx 0$ represents the accumulated null observations and $\sigma_2$ quantifies their observational imprecision. With $\sigma_2 = 5$, the posterior distribution of $\theta$ given $w = 0$ becomes

$$\theta \mid w, H_2 \sim \mathcal{N}(0, \tau_{\text{post}}^2),$$

with

$$\frac{1}{\tau_{\text{post}}^2} = \frac{1}{\tau^2} + \frac{1}{\sigma_2^2}.$$

For the numerical values $\tau = 20$ and $\sigma_2 = 5$, this yields $\tau_{\text{post}} \approx 4.85$. The updated Newtonian model therefore becomes

$$y \mid H_2, w \sim \mathcal{N}(0, \sigma^2 + \tau_{\text{post}}^2),$$

with predictive standard deviation approximately 4.88. Under this distribution, the probability density at $y = 43$ is

$$p(y \mid H_2, w) = \frac{1}{\sqrt{2\pi(\sigma^2 + \tau_{\text{post}}^2)}} \exp\left(-\frac{43^2}{2(\sigma^2 + \tau_{\text{post}}^2)}\right),$$

which evaluates to approximately $1.3 \times 10^{-18}$. At this point, the Newtonian theory supplemented by the Vulcan hypothesis assigns virtually zero integrated probability to the actual data.

By contrast, Einstein's general relativity predicts the anomalous precession directly and without adjustable parameters. The predicted additional precession is

$$\mu_{\text{GR}} = \frac{6\pi GM}{a(1-e^2)c^2},$$

where $a$ is the semi–major axis of Mercury's orbit, $e$ its eccentricity, $M$ the mass of the Sun and $c$ the speed of light. Substituting astronomical constants yields a value of approximately forty–three arcseconds per century. Under the relativistic model $H_3$, therefore,

$$y \mid H_3 \sim \mathcal{N}(\mu_{\text{GR}}, \sigma^2),$$

and at $y = 43$ the likelihood is

$$p(y \mid H_3) = \frac{1}{\sqrt{2\pi\sigma^2}} \approx 0.798.$$

The Bayes factor comparing general relativity to the Newton–plus–Vulcan model is therefore

$$B_{32} = \frac{p(y \mid H_3)}{p(y \mid H_2, w)},$$

and with the numerical values above one obtains

$$B_{32} \approx \frac{0.798}{1.3 \times 10^{-18}} \approx 6 \times 10^{17}.$$

The posterior odds in favour of Einstein's theory are overwhelming regardless of the prior odds, unless one assigns astronomically implausible priors to Newtonian mechanics relative to general relativity.

This analysis shows that the anomaly in Mercury's orbit does not merely count against Newtonian gravity in a logical sense. Rather, it destroys the Newtonian model because the best auxiliary hypothesis available within the Newtonian framework assigns vanishing marginal likelihood to the observed data once all constraints are accounted for. At the same time, the rival theory of general relativity assigns extremely high probability to the anomaly and does so without resorting to adjustable parameters. In a Bayesian sense, Newtonian gravity is eliminated not because it is logically refuted but because its posterior probability becomes negligibly small relative to that of general relativity.

The contrast with the Neptune case is striking. There, Newtonian mechanics survived because its predictive adequacy could be restored by introducing a parsimonious and physically well–motivated auxiliary hypothesis whose presence dramatically increased the theory's marginal likelihood. Here, the same strategy fails. The available auxiliary hypotheses are inconsistent with auxiliary evidence, and their integration into the model lowers rather than increases the marginal likelihood. The Newtonian framework therefore degenerates in the Lakatosian sense, while Einstein's theory emerges as a progressive rival with far superior explicativity. The Mercury anomaly thus provides a clear example of eliminative inference: a theory is abandoned when no permissible modification within its own framework allows it to retain competitive posterior support.

## 7. Falsification as Bayesian Elimination: A General Framework

The two case studies considered above suggest a general reinterpretation of the concept of falsification. In its classical formulation, falsification is presented as a logical relation between a theory and an observation that contradicts its predictions. This picture presupposes that each theory makes deterministic claims about the world, and that the role of evidence is to determine whether these claims are true or false. The philosophical difficulties associated with this view are well known. The Duhem–Quine thesis emphasises that predictions depend on auxiliary assumptions, so that a failed prediction can always be attributed to an auxiliary hypothesis rather than to the core theory. Popper attempted to mitigate this difficulty by insisting that scientific testing is comparative and that a theory is rejected only when a superior rival is available. Yet his account remained qualitative because it did not specify how comparative superiority should be measured.

The Bayesian framework resolves these difficulties by shifting the focus from logical relations to probabilistic ones. A scientific theory is represented not as a set of deterministic claims about observable outcomes but as a probability measure over the space of possible data. The question of whether a theory is refuted becomes the question of whether it assigns sufficiently high probability to the actual observations when evaluated in comparison with rival theories. Let $M_1, \ldots, M_k$ denote a set of competing models and let $D$ denote the observed data. The posterior probability of model $M_i$ is

$$p(M_i \mid D) = \frac{p(M_i)p(D \mid M_i)}{\sum_{j=1}^{k} p(M_j)p(D \mid M_j)},$$

where $p(M_i)$ is the prior probability of the model and $p(D \mid M_i)$ its marginal likelihood. Falsification, in this setting, corresponds to the situation in which the posterior probability of a model becomes negligibly small compared with that of at least one alternative. The elimination of a theory is therefore relative, not absolute, and depends on the comparative predictive success of the competing models.

Within this framework, the concept of refutation appears as a limit case of posterior collapse. Suppose two models $M_1$ and $M_2$ are assigned equal prior probabilities. Their posterior odds are then

$$\frac{p(M_1 \mid D)}{p(M_2 \mid D)} = \frac{p(D \mid M_1)}{p(D \mid M_2)} = B_{12}.$$

If $B_{12}$ is extremely small, the posterior probability of $M_1$ becomes negligible. In this sense, the theory $M_1$ is eliminated by the data, but the elimination is not a logical contradiction. It arises because the events that $M_1$ assigns high probability to are not the events that were actually observed. Scientific refutation, on this view, is a probabilistic judgement about the explanatory adequacy of competing models rather than a deductive relation between theory and observation.

One consequence of this perspective is that the role of auxiliary hypotheses is clarified rather than obscured. A theory together with its auxiliaries forms a composite model whose plausibility is determined by its marginal likelihood. If the auxiliaries can be adjusted in a way that increases the marginal likelihood, then the composite model gains credibility. If, however, the auxiliaries must be finely tuned to accommodate the data, their contribution to the marginal likelihood becomes negative, because the parameter volume over which the likelihood is high becomes small relative to the prior volume. The theory then loses plausibility. In this way, the Bayesian framework internalises Occam's razor and provides a natural explanation for why ad hoc hypotheses are penalised. They reduce the integrated probability that the model assigns to the observable world.

This provides a quantitative foundation for Lakatos' distinction between progressive and degenerative research programmes. A programme is progressive when modifications to its auxiliary assumptions increase the marginal likelihood by predicting new observations or by explaining anomalies with minimal additional complexity. A programme is degenerative when modifications decrease the marginal likelihood by introducing fine–tuned adjustments that increase complexity without offering genuine predictive improvements. Bayesian model comparison formalises this distinction by assigning posterior probability to entire theoretical structures rather than to isolated hypotheses.

Popper's emphasis on severe tests also finds a natural expression in this framework. A test is severe when the competing models assign sharply different probabilities to the relevant observations. If both models predict the data with similar marginal likelihoods, then the test is uninformative. If one model assigns a significantly higher probability to the data than its rival, then the test is severe in the sense that it yields a decisive shift in posterior odds. The notion of severity therefore becomes the statement that the likelihood ratio is large or small, depending on which theory is favoured. This is not a deviation from Popper's original intent but rather its mathematical completion.

The Bayesian reinterpretation of falsification also clarifies the distinction between anomaly and refutation. An anomaly is simply an observation that has low probability under a model. Such an observation does not automatically lead to refutation. It gains significance only when another model assigns a higher probability to the same observation. This is precisely what occurred in the case of Mercury's precession. The anomaly was significant not because it contradicted Newtonian mechanics directly but because general relativity predicted it exactly and without free parameters. In contrast, the anomaly in Uranus' orbit did not threaten Newtonian gravity because the planetary hypothesis offered a superior explanation within the same theoretical framework. In both cases, the decisive factor was the comparative marginal likelihood, not the mere presence or absence of discrepancies.

The general framework that emerges is therefore eliminative rather than confirmatory. Theories are not proved by data; they survive by outperforming their rivals in predictive accuracy and explicativity. Theories are not logically falsified by data; they are abandoned when their marginal likelihood collapses relative to that of alternative models. This interpretation reconciles Popper's insistence on critical comparison with the Bayesian emphasis on posterior probabilities. It also clarifies why scientific progress is not a sequence of confirmations but a process of successive eliminations in which only the most predictive and parsimonious theories remain competitive.

## 8. Discussion

The reinterpretation of falsification as a Bayesian process of model elimination sheds light on several longstanding debates in the philosophy of science. The historical tension between Popper's critical rationalism and Bayesian epistemology has often been portrayed as a fundamental incompatibility. Popper rejected probabilistic accounts of scientific inference on the grounds that they conflated evidential support with psychological belief, while Bayesian theorists dismissed Popper's falsificationism as an idealised account that could not accommodate the actual practice of scientific reasoning. The analysis developed in this paper suggests that this opposition is misplaced. Once falsification is understood as a comparative rather than an absolute notion, its structure becomes not only compatible with Bayesian reasoning but naturally expressible within it.

One central theme clarified by the Bayesian framework is the dual role of prediction and explanation in scientific theory choice. Popper emphasised prediction as the hallmark of a good theory and held that a theory earns corroboration only when it successfully passes tests that it could have failed. Bayesian inference accommodates this principle by assigning higher posterior probability to models that concentrate their prior mass on regions of the sample space consistent with the observed data. In this sense, predictive adequacy is encoded in the marginal likelihood. At the same time, Good's notion of explicativity introduces a complementary dimension: a theory is preferable if it captures the structural relations among the data without requiring extensive fine–tuning. The Neptune case demonstrates this clearly. The auxiliary hypothesis of an unseen planet increased both the predictive power and explicativity of the Newtonian model, thereby raising its marginal likelihood and posterior probability. The Mercury case shows the opposite pattern: the auxiliary hypothesis required substantial fine–tuning and was incompatible with auxiliary evidence, resulting in a sharp reduction in marginal likelihood.

This duality between prediction and explanation offers a resolution to the apparent conflict between Popper's emphasis on severe tests and Jaynes' emphasis on explanatory power. Popperian severity concerns the ability of a test to discriminate between competing theories. From a Bayesian perspective, severity is the condition that competing models assign substantially different probabilities to the same data. Jaynes' explicativity concerns the concentration of probability mass on the structural features of the data. These two notions converge in the behaviour of the marginal likelihood, which rewards models that anticipate the observed data with high integrated probability and penalises models that rely on adjustments that increase complexity without improving prediction.

The Bayesian reinterpretation also illuminates the logic of scientific revolutions. Thomas Kuhn argued that scientific change proceeds through periods of normal science punctuated by crises in which anomalies accumulate until a paradigm shift occurs. Kuhn's language is qualitative, but the Bayesian framework provides a quantitative account of the phenomenon he described. Normal science corresponds to periods in which the dominant model retains high posterior probability because its marginal likelihood remains competitive with that of its rivals. A crisis occurs when the cumulative weight of evidence reduces the model's marginal likelihood to the point where alternatives command greater posterior support. The transition to a new paradigm corresponds to a posterior redistribution in which the previously dominant model becomes negligible. Kuhn's concept of incommensurability appears less problematic when viewed from this perspective, because models can be compared as probabilistic structures even when they differ in their conceptual vocabulary.

Another important implication of the Bayesian view concerns the status of ad hoc hypotheses. Popper criticised ad hoc modifications as attempts to immunise a theory against refutation by adjusting its auxiliary assumptions without increasing its testable content. Bayesian analysis explains why such manoeuvres are epistemically illegitimate. A modification that reduces the domain of prediction or requires highly specific parameter values shrinks the region of parameter space in which the model assigns high likelihood to the data. Since the marginal likelihood integrates the likelihood over the full parameter space, such modifications reduce the model's overall predictive probability, even if the fitted likelihood increases at a single point. The Bayesian penalty for ad hocness is therefore not imposed externally but arises automatically from probabilistic coherence.

Finally, the Bayesian reinterpretation of falsification avoids the paradoxes associated with the classical view of logical refutation. In the traditional picture, a single anomalous observation can discredit a theory, but this leads to an impractical view of scientific practice in which theories would be abandoned prematurely. In the Bayesian view, anomalies have evidential force only relative to the success of alternative explanations. A theory is abandoned not because it encounters a recalcitrant datum but because the totality of evidence makes a rival theory overwhelmingly more plausible. This corresponds closely to Popper's mature view, in which theory rejection occurs only when a superior alternative is available. Bayesian inference supplies the quantitative mechanism that Popper lacked, grounding eliminative induction in the mathematics of posterior probability.

The overall picture that emerges is one in which falsification is not a dramatic event but a gradual and cumulative process. Theories lose or gain plausibility as new data are incorporated and as the comparative structure of evidential support shifts. Scientific progress becomes a dynamic reallocation of posterior probability within a space of competing models. This perspective not only harmonises Popper's and Jaynes' views but also clarifies why scientific inquiry is both fallible and rational. It is fallible because no theory is ever conclusively verified or refuted; it is rational because the principles of Bayesian model comparison provide a coherent guide to the revision of belief in the light of evidence.